\documentclass{article}

\usepackage{arxiv}

\usepackage[utf8]{inputenc} % allow utf-8 input
\usepackage[T1]{fontenc}    % use 8-bit T1 fonts
\usepackage{hyperref}       % hyperlinks
\usepackage{url}            % simple URL typesetting
\usepackage{booktabs}       % professional-quality tables
\usepackage{amsfonts}       % blackboard math symbols
\usepackage{nicefrac}       % compact symbols for 1/2, etc.
\usepackage{microtype}      % microtypography
\usepackage{lipsum}

\usepackage{amsmath,amssymb,amsfonts}
\usepackage{algorithmic}
\usepackage{graphicx}
\usepackage{textcomp}
\usepackage{xcolor}
\newtheorem{attackscenario}{Attack Type}
\DeclareMathOperator*{\median}{median}
\usepackage[font=footnotesize]{subfig}
\usepackage{todonotes}
\usepackage{tikz}
\usetikzlibrary{arrows,positioning} 
\usetikzlibrary{fit}

\usepackage[normalem]{ulem}
\newcommand{\blue}[1]{\textcolor{black}{#1}}

\tikzset{
	>=stealth',
	node/.style={
		rectangle,
		rounded corners,
		draw=black, very thick,
		minimum height=2em,
		text centered},
	data/.style={
		rectangle,
		draw=black, very thick,
		minimum height=2em,
		text centered},
	arrow/.style={
		->,
		thick,
		shorten <=2pt,
		shorten >=2pt,}
}

\title{Monte Carlo and Reconstruction Membership Inference Attacks against Generative Models}

\author{
  Benjamin Hilprecht\thanks{A part of this work was done during an internship at SAP SE.} \\
  TU Darmstadt\\
  Darmstadt, Germany \\
  \texttt{benjamin.hilprecht@cs.tu-darmstadt.de} \\
   \And
 Martin H\"{a}rterich, Daniel Bernau \\
  SAP SE\\
  Karlsruhe, Germany\\
  \texttt{firstname.lastname@sap.com} \\
}

\begin{document}
\maketitle

\begin{abstract}
{% !TeX root = mi_attack_against_gan.tex
We present two information leakage attacks that outperform previous work on membership inference against generative models. 
The first attack allows membership inference without assumptions on the type of the generative model. Contrary to previous evaluation metrics for generative models, like Kernel Density Estimation, it only considers samples of the model which are close to training data records. 
The second attack specifically targets Variational Autoencoders, achieving high membership inference accuracy. 
Furthermore, previous work mostly considers membership inference adversaries who perform single record membership inference. We argue for considering regulatory actors who perform set membership inference to identify the use of specific datasets for training. The attacks are evaluated on two generative model architectures, Generative Adversarial Networks (GANs) and Variational Autoencoders (VAEs), trained on standard image datasets. Our results show that the two attacks yield success rates superior to previous work on most data sets while at the same time having only very mild assumptions. We envision the two attacks in combination with the membership inference attack type formalization as especially useful. For example, to enforce data privacy standards and automatically assessing model quality in machine learning as a service setups. In practice, our work motivates the use of GANs since they prove less vulnerable against information leakage attacks while producing detailed samples.}
\end{abstract}

% keywords can be removed
\keywords{Machine Learning, Privacy}
% !TeX root = mi_attack_against_gan.tex
\section{Introduction}
\label{sec:intro}
%%Roadmap:
% ML
% MI Problem und relevanz bei GANs
% Previous MI Attack scenarios & issues
Machine learning is ubiquitous in software applications nowadays. However, the success of machine learning (ML) depends as much on sophisticated algorithms as it does on the availability of large sets of training data.
Gathering sufficient amounts of training data for satisfying model generalization has proven cumbersome especially for sensitive data and, in some cases, resulted in privacy violations due to data misuse (e.g., the inappropriate legal basis for the use of National Health Service (NHS) data in the DeepMind project~\cite{Guardian2017NHS, SkyNews2017NHS, Hayes2019}).
The desire to identify on which data a model was trained, and thus detect privacy violations gave rise to model inversion, which aims for reconstructing a training dataset with missing parts \cite{fredrikson2015model}, and membership inference~(MI)~\cite{shokri2017membership}. Within this work we address the latter, striving to identify whether an individual or a set of individuals, {\em belong to} a certain training dataset.
 
\blue{Motivated by the recent NHS misuse case we consider two membership inference actors: an adversary performing {\em single} record MI and a regulator performing {\em set} MI. Single MI is used in previous work to model an adversary who is mainly interested in identifying individuals within a dataset. However, set MI is relevant for regulatory audits since it can be used to prove that a specific set of records was used to train a model. If the practitioner who trained the model was not authorized to use a specific dataset for this purpose regulators can apply set MI to prove data privacy violations.}

We propose and evaluate two novel membership inference attacks against recent generative models, \textit{Generative Adversarial Networks}~(GAN)~\cite{goodfellow2014generative} and \textit{Variational Autoencoders}~(VAE)~\cite{kingma2013auto}.
These generative models have become effective tools for (unsupervised) learning with the goal to produce samples of a given distribution after training. Generative models thus have many applications like the synthesis of photo-realistic images, image-to-image translation, and even text~\cite{bowman2015generating} or sound~\cite{donahue2018synthesizing} synthesis. 
However, the MI attack of Shokri et al.~\cite{shokri2017membership} against discriminative models is not directly applicable to generative models and thus alternative means are required. Moreover, previous attacks on generative models were specialized on GANs~\cite{Hayes2019}. In contrast, our first attack is applicable to every generative model from which one can draw samples. The attack only considers samples which are very close to train or test records giving it an edge over existing methods like the Euclidean attack \cite{hayes2017logan}.
The second proposed attack is solely applicable to Variational Autoencoders. Hence, our attacks allow membership inference attacks against a broader class of generative models. In some cases, the attacks formulated in this work yield accuracies close to $100\%$, clearly outperforming previous work. Furthermore, the regulatory actor performing set MI helps to unveil even slight information leakage. Hence, set MI is of high practical relevance for enforcing data privacy standards.

The close connection of information leakage to overfitting provides another motivation for this work. We intuitively relate overfitting to memorization of training data, since strong overfitting will result in the replication of given data in generative models and therefore higher accuracies of membership inference attacks. Given that in extreme cases a linear relationship between the success of membership inference attacks and overfitting has been observed for discriminative models~\cite{fredrikson2015model} we also want to avoid overfitting in the case of generative models. However, overfitting is neither straightforward to define nor identify for generative models. 

As proposed by Hayes et al.~\cite{Hayes2019}, the accuracy of attacks in single MI can be used as an indicator for overfitting. We thoroughly compare our attacks against state of the art attacks on generative models introduced by Hayes et al.~\cite{Hayes2019} to further investigate this claim. The proposed type of set membership inference results in higher accuracy values and is potentially a means for identifying even slight overfitting in generative models. For machine learning as a service (MLaaS) our attacks are therefore potentially a means for automatically assessing the quality of the learned generative model more accurately than previous approaches. %Daniel: Ich würde hier auf den Zeilenumbruch verzichten zu gunsten einer einfacheren Struktur. Durch die Liste wird die Abgrenzung ja implizit aufgezeigt.
The main contributions of this work are:
\begin{itemize}
\item
a membership inference attack based on Monte Carlo integration that exclusively considers small distance samples from the model,
%Two variants of a novel %class of 
%membership inference attack based on Monte Carlo Integration which can be applied to arbitrary generative models. \blue{A key insight is that these attacks only work if exclusively very close samples are considered.}
\item
a membership inference attack designed for Variational Autoencoders: the \textit{Reconstruction attack},
%A novel membership inference attack specifically against Variational Autoencoders yielding high accuracies.
\item
and a membership inference variation performing \textit{set membership inference}, which is systematically evaluated and which we envision to be used by regulators to enforce data privacy standards. 
\end{itemize}

We evaluated the attacks on the image datasets MNIST, Fashion-MNIST, and CIFAR-10 for both Generative Adversarial Networks (GANs) and Variational Autoencoders (VAEs) which are widely used generative models. For VAEs, the Reconstruction attack yielded accuracies close to 100\% for set MI and between 57\% and 99\% for single MI. The MC attack reached between 72\% and 100\% set MI accuracy and up to 60\% accuracy for single MI. The attacks were less effective on GANs in our experiments. However, the MC attack accuracies against GANs range from 65\% to 75\% for set MI. In general, the MC attack performs better if the samples drawn from the model are of high quality.

This paper is structured as follows.
Section~\ref{sec:background} explains the threat model and membership inference attacks considered in this work. In particular, we introduce and formalize two actors who perform single and set membership inference. Furthermore, we argue for their relevance in real-world use cases.
In Section~\ref{sec:novel} we introduce and formalize our two attacks which are applicable to both single and set membership inference. To this end, details regarding GANs and VAEs are provided.
The subsequent Section~\ref{sec:experiments} contains an evaluation of our attacks on reference datasets.
Related work is discussed in Section~\ref{sec:related}.
A summary and outlook (Section~\ref{sec:conclusion}) concludes the paper.

% !TeX root = mi_attack_against_gan.tex
% Better title needed
\section{Membership Inference Attacks}
\label{sec:background}
In this section, we introduce the threat model and the two kinds of attacks considered in this paper: single MI and set MI.
We start the section by exposing some background on MI.

\subsection{Background of Membership Inference}
\label{sec:prel_memb}

The goal of membership inference (MI) is to gather evidence whether a specific record or a set of records belongs to the training dataset of a given machine learning model. MI thus represents an approach for measuring how much a model leaks about individual records of a population. %beyond what it reveals about an arbitrary member of the population. 
The success rates of MI attacks against a model are tightly linked to overfitting (i.e., the generalization error \cite{yeom2017unintended}). The poorer a model generalizes the more specificities it contains about individual training data records.

In this work, two kinds of MI are considered: single MI and set MI. The single MI is comparable to common experiment setups for MI\cite{shokri2017membership,Hayes2019}. In the set MI setting a regulator has to recognize which of the two provided sets contains training data records. 
%Before we provide details and formalize the attacks, we introduce our threat model.

\subsection{Threat Model}
\label{sec:attack_scenarios}

\begin{table*}
	\centering
	\caption{Comparison of Attacks}
	\label{tab:comp_attacks}
	\begin{tabular}{lp{4.2cm}p{2.4cm}p{6cm}}
		\hline
		Attack & Required Access & Applicable & Idea\\
		\hline
		White-box & Discriminator & GANs & Evaluate Discriminator\\
		Black-box & Samples from Generative Model & Generative Models & Train auxiliary GAN on samples and evaluate Discriminator\\
		Monte Carlo & Samples from Generative Model & Generative Models & Monte Carlo approximation on close samples\\
		Reconstruction Attack & VAE model & VAEs & VAE reconstructs training data more precisely\\
		\hline
	\end{tabular}
\end{table*}
This work considers two actors corresponding to single and set MI, respectively. The first actor is an honest-but-curious adversary $\mathcal{A}$ and the second actor is a regulatory body $\mathcal{R}$. Each actor focuses on a specific task: adversary $\mathcal{A}$ is common in MI literature and engages in a \textit{single} membership inference to infer whether a single record known to him was present in the training dataset of the target model. The regulatory body $\mathcal{R}$ performs \textit{set} membership inference to identify whether a set of records was present in the training dataset. This attack can provide evidence that a certain set of training data was illegally used to train a generative model.

Both actors are assumed to have no access to the underlying training dataset of the generative model, and they refrain from activities that maliciously modify this target model.
The actors $\mathcal{A}$ and $\mathcal{R}$ can both launch the Monte Carlo (MC) attack as well as the Reconstruction attack. (See Section~\ref{sec:novel} for details.) The choice of the attack determines the requirements on the information that is available to the actor. The MC attack requires samples drawn from the generative model while the Reconstruction attack has to be able to evaluate the generative model.

\subsection{Adversarial Actor: Single MI}

Single MI has been used by previous work to evaluate attacks against GANs~\cite{Hayes2019}. In this setting, the honest-but-curious adversary $\mathcal{A}$ has to identify individual records which were used to train the model.
To this end $M$ records from the training data and $M$ records from the test dataset $\{x_1,\dots,x_{2M}\}$ are given. Both the MC attack and the Reconstruction attack rely on a function $\hat{f}(x)$ that can be computed for each of the records. The intuition is that this function attains higher values for training data records. Details on how this function is realized are given in the next section. In the following description of the attack types we use the general notation $\hat{f}(x)$. 

For every record $x_{i},$ $\mathcal{A}$ has to decide whether it was part of the training data. In general, $\mathcal{A}$ picks the $M$ records with the $M$ greatest values of the function $\hat{f}(x).$

\begin{attackscenario}[Single Membership Inference]
Let $\mathcal{A}$ be an adversary who is able to compute the function $\hat{f}(x)$ for every record $x$.
\begin{enumerate}
	\item Choose records $\{x_{1},\dots,x_{M}\}$ from the training data. 
	\item Choose records $\{x_{M+1},\dots,x_{2M}\}$ from the test data. 
	\item $\mathcal{A}$ is presented the set $\{x_{1},\dots,x_{2M}\}.$
	\item $\mathcal{A}$ labels the $M$ records with highest values $\hat{f}(x_i)$ as training data.
\end{enumerate}
\end{attackscenario}

We denote the $M$ records chosen by $\mathcal{A}$ as $\{x^{\mathcal{A}}_1,\dots,x^{\mathcal{A}}_M\}.$ We call the proportion of actual training data in this set
\begin{equation*}
\frac{1}{M}\cdot\left|\{i\;|\;x^{\mathcal{A}}_i\in\{x_{1},\dots,x_{M}\}\}\right|
\end{equation*}
the accuracy of the attack for single MI.
%We define accuracy of the attack as the average proportion of correctly identified training records by $\mathcal{A}$. For one experiment the accuracy amounts to
%\begin{equation}
%\frac{1}{M}\sum_{i=1}^{M}
%\end{equation}

\subsection{Regulatory Actor: Set MI}
\label{subsec:set-mi}

Set MI corresponds to the needs of regulators and auditors aiming to prove data privacy violations in machine learning.
One set consisting of $M$ records from the training data $\{x_1,\dots,x_{M}\}$ and another set consisting of $M$ records from the test data $\{x_{M+1},\dots,x_{2M}\}$ are shown to a regulator $\mathcal{R}$ in either order. The task of $\mathcal{R}$ is to decide which of the two sets is a subset of the original training data. Contrary to single MI, $\mathcal{R}$ knows which records belong to the same data source (training data or test data). However, $\mathcal{R}$ does not know which set is a subset of the original training data.

Similar to single MI $\mathcal{R}$ computes the function $\hat{f}(x)$ for every record and selects the $M$ records with the $M$ highest values $\hat{f}(x)$. For each of the selected records, $\mathcal{R}$ checks to which set it belongs and eventually selects the set from which most of these records stem as subset of the original training data.%
\footnote{If an equal number of records belong to the first and the second set, $\mathcal{R}$ picks one of the sets with probability $50\%.$}
Note that this is equivalent to taking the set with the higher median. Since we do not have any prior knowledge on the type of distribution of the $\hat{f}$-values this is more robust than considering e.g.\ the mean.

\begin{attackscenario}[Set Membership Inference]
Let $\mathcal{R}$ be an adversary able to calculate the function $\hat{f}(x)$ for every record $x$.	
	\begin{enumerate}
		\item Choose records $\{x_{1},\dots,x_{M}\}$ from the training data. 
		\item Choose records $\{x_{M+1},\dots,x_{2M}\}$ from the test data.
		\item $\mathcal{R}$ is presented the sets $\{x_{1},\dots,x_{M}\}$ and $\{x_{M+1},\dots,x_{2M}\}.$
		\item $\mathcal{R}$ identifies the $M$ records with highest values $\hat{f}(x_i).$
		\item $\mathcal{R}$ chooses the set from which most of these records stem.
		\item If both have the same number of representatives $\mathcal{R}$ picks one set randomly.
	\end{enumerate}
\end{attackscenario}

The accuracy of an attack of this type is defined as the average success rate of $\mathcal{R}$, i.e., the probability that $\mathcal{R}$ identifies the true subset of the training data. 

\subsection{Relevance for Real-World Use Cases}
\label{sec:application}

The formalized MI attack types are an alternative to assessing a single record $x$ by computing $\hat{f}(x)$ and considering the record part of the training data if the value exceeds a threshold. %For the attack evaluation false positive and false negative rates for thresholds of different attacks could be compared. 
While the single record approach is conceptually similar, the formalized types contributed in this work are closer to real-world use cases. 
For example, in machine learning as a service (MLaaS) applications access to both test and training data is implicitly given. Hence, the single MI and set MI attack types can be automatically conducted. High MI attack accuracies suggest that the model quality is insufficient w.r.t.~privacy.

\begin{figure}[!t]
	\centering
	\resizebox{\columnwidth}{!}{
		\begin{tikzpicture}
		\path 
		(0,0) coordinate (A) ellipse (2 and 1.5) [draw]
		(-0.4,0) ellipse (1.5 and 1.1) [draw]
		(-0.8,0) ellipse (1 and 0.75) [draw]
		(-1.2,0) ellipse (0.5 and 0.38) [draw]
		(3.5,0) ellipse (1 and 0.75) [draw]
		(3.1,0) ellipse (0.5 and 0.38) [draw]
		(-1.2,0) -- (5,3.5) -- (5.5,3.5) node[right]{train set of regulator's MI attack (a)}
		(-0.3,0) -- (5,3) -- (5.5,3) node[right]{suspected illegal use (b)}
		(0.6,0) -- (5,2.5) -- (5.5,2.5) node[right]{actual illegal use (c)}
		(1.5,0) -- (5,2) -- (5.5,2) node[right]{complete training data used (d)}
		(3.1,0) -- (5,1.5) -- (5.5,1.5) node[right]{test set of regulator's MI attack (e)}
		(4,0) -- (5,1) -- (5.5,1) node[right]{test data of regulator (f)};
		\end{tikzpicture}
	}
	\caption{Venn diagram of training and test data in the regulatory use case for $\mathcal{R}$.}
	
	\label{fig:venn}
\end{figure}
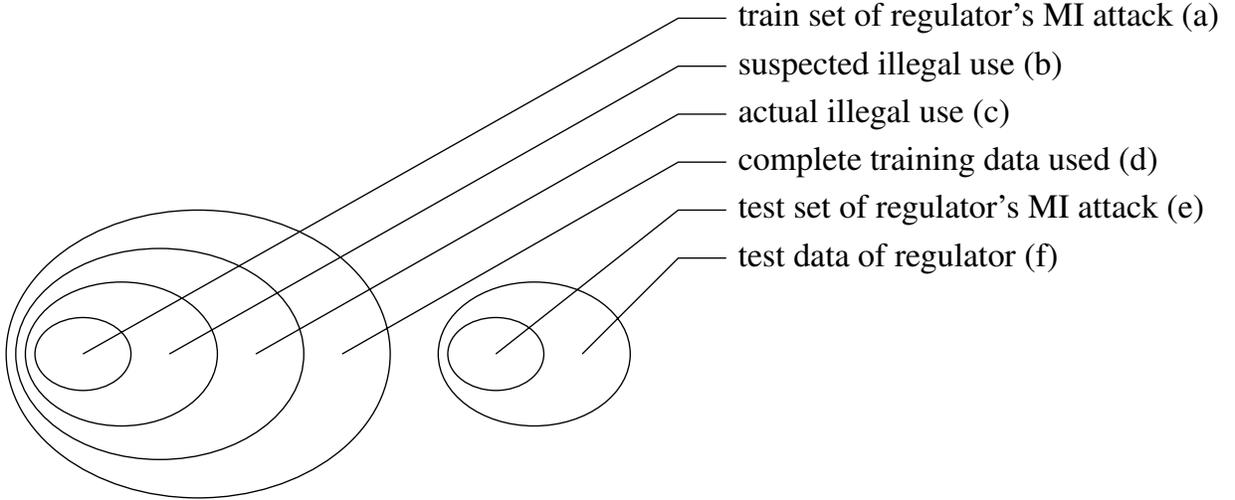

Figure \ref{fig:venn} visualizes the regulatory use case. The regulator $\mathcal{R}$ suspects that a certain dataset was illegally used to train a model (b). Actually, even more data was used illegally (c). Moreover, some legally obtained data might have been used. Together with the illegal data, it represents the complete training data (d).
$\mathcal{R}$'s set of suspected data is used as train set in the set MI attack (a). $\mathcal{R}$ also needs test data (f) from which a subset (e) is used as test set for the attack. If the attack is successful the illegal use can be proven. Otherwise, the attack does not perform better than random guessing. By repeating the attack for multiple choices of subsets (a) and (f) $\mathcal{R}$ ensures statistical significance. Note that $\mathcal{R}$ does not need to know the entire training data since the MI attacks also work for subsets of the entire training data. The accuracy does not depend on the concrete subset choice as we will show in our experiments in Section~\ref{sec:experiments}. 

Note that in both single and set MI we assume that there are exactly as many test as train records. In the regulatory use case of set MI this is realistic since a sample of the larger of the two sets can be used if they are not of equal size. To make the results of single and set MI comparable, and to be in line with the balanced setting in previous work~\cite{shokri2017membership}, we also decided to use this setup in single MI. Note that this is potentially an advantage for~$\mathcal{A}$.

% !TeX root = mi_attack_against_gan.tex

%\section{MI Attack types}

% !TeX root = mi_attack_against_gan.tex

\section{Attack Details}
\label{sec:novel}
\begin{figure*}[!th]
	\centering
	\begin{tikzpicture}[node distance=1cm, auto,]
	
	\node[data] (dx) {$D(x)$};
	
	\node[node, left=of dx] (dis) {Discriminator $D$}
	edge[arrow] (dx.west);
	
	\node[data, above left=1cm of dis] (fake) {$G(z)$}
	edge[arrow] (dis.west);
	
	\node[node, left=of fake] (gen) {Generator $G$}
	edge[arrow] (fake.west);
	
	\node[data, left=of gen] (z) {$z\sim p_{noise}$}
	edge[arrow] (gen.west);
	
	\node[data, below left=1cm of dis] (train) {Training Data}
	edge[arrow] (dis.west);
	
	\end{tikzpicture}
	\caption{Architecture of a Generative Adversarial Network (GAN).}
	\label{fig:gan_arch}
\end{figure*}
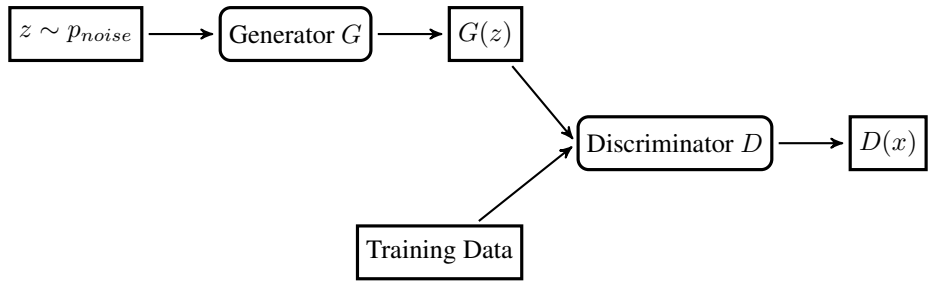
In this section we introduce two novel MI attacks. They can be used for both single and set MI. The first attack, namely the \textit{Monte Carlo} attack~(Section~\ref{subsec:monte_carlo}) compares samples drawn from the model to either test or train records. Opposing to existing approaches, only very close samples are considered. Indeed, this distinguishes the attacks from previous approaches like the Euclidean attack \cite{hayes2017logan} and made the attacks effective. Furthermore, the \textit{Reconstruction} attack~(Section~\ref{subsec:rec_attack}) which is optimized for VAEs is presented. A comparison of our attacks and state-of-the-art attacks is given in Table~\ref{tab:comp_attacks}. Again, an attack is fully specified by the function $\hat{f}(x)$ which will be introduced in the following. Since in the description of the attacks details about generative models are required, we briefly describe VAEs and GANs in the next section.

\subsection{Generative Models}
\label{sec:prel_gen}

Generative models are ML models that are trained to learn the joint probability distribution $p(X,Y)$ of features $X$ and labels $Y$ of training data.
In this paper we apply two decoder based models relying on neural networks, namely \textit{Generative Adversarial Networks}~(GANs) \cite{goodfellow2014generative} and \textit{Variational Autoencoders}~(VAEs)~\cite{kingma2013auto}. Note, however, that our Monte Carlo attack is applicable to all generative models from which one can draw samples. The reconstruction attack specifically targets VAEs.
\subsubsection{Generative Adversarial Networks}
\label{sec:gan}
A GAN %\cite{goodfellow2014generative} is a generative model which is trained in an adversarial manner. It 
consists of two competing models, a \textit{generator} $G$ and a \textit{discriminator} $D$, which are trained in an adversarial manner (i.e., compete against each other).
We describe the approach in detail referring to Figure~\ref{fig:gan_arch}. 

%The generator $G$ of the GAN provides input to the generator network by sampling data $z$ from a prior distribution $p_{noise}$ (e.g., Gaussian). 
To generate artificial data a prior $z$ is sampled from a prior distribution $p_{noise}$ (e.g., Gaussian) and fed as input into the generator $G.$
%Discriminator $D$ outputs the probability that a given sample stems either from the training data or $G$. While 
The task of the discriminator $D$ is to output the probability that generated samples stem either from the training data or $G.$ However, $G$ tries to fool $D$ by generating samples that $D$ misclassifies. Hence, the outputs $G(z)$ should look similar to the training data $x$ (i.e.~records sampled from $p_{data}$).
This is expressed as a \blue{two-player} zero-sum game via the following objective function:
\begin{equation*}
\min_{G} \max_{D} \mathbb{E}_{x\sim p_{data}}[\log D(x)]+ \mathbb{E}_{z\sim p_{noise}}[\log(1-D(G(z)))].
\end{equation*}

Gradients are computed for $G$ and $D$ during training, and usually, after already a few steps of training $G$ produces realistic outputs. %\red{In a large-scale study, it was shown that the original GAN is still competitive against suggested modifications \cite{lucic2017gans}. Hence, we examine only the original architecture within this work.}
A conditional generative model is obtained by providing a condition $c$ (e.g., a class label) as an input both to the generator and the  discriminator~\cite{goodfellow2014generative}.

\subsubsection{Variational Autoencoders}
\label{sec:vae}

VAEs~\cite{kingma2013auto} consist of two networks - an encoder $E$ and a decoder $D$. During training each record $x$ is given to the encoder which outputs the mean $E_\mu(x)$ and variance $E_\Sigma(x)$ of a Gaussian distribution. A latent variable $z$ is sampled from this distribution $N(E_\mu(x),E_\Sigma(x))$ and fed into the decoder $D$. The reconstruction $D(z)$ should be close to the training data record $x$. 

During training two terms need to be minimized. First, the reconstruction error $\lVert D(z)-x \rVert.$ Second $\textit{KL}(N(E_\mu(x),E_\Sigma(x)) || N(0,1)),$ the \textit{Kullback-Leibler divergence} between the distribution of the latent variables $z$ and the unit Gaussian. The second term prevents the network from only memorizing certain latent variables because the distribution should be similar to the unit Gaussian. In practice, both the encoder $E$ and the decoder $D$ are neural networks. Kingma et al.~\cite{kingma2013auto} provide details on how to train those networks given the training objective with the reparametrization trick. Moreover, they motivate the training objective as a lower bound on the log-likelihood.
Sampling from the VAE is achieved by sampling a latent variable $z\sim N(0,1)$ and passing $z$ through the decoder network $D$. The outputs of the decoder $D(z)$ then serve as samples. Like for GANs, a conditional variant is obtained by providing a condition $c$ as input to the decoder and the encoder. 

% ELBO lower bound to log-likelihood. 
% 
% VAEs~\cite{kingma2013auto} are based on the underlying assumption that the data is generated by a two step random process involving a latent variable $z$. 
% First, a value $z$ is sampled from a prior distribution $p(z)=p_{noise}$. Second, the data $x$ is sampled from a conditional distribution $p(x|z)$. This distribution is usually a Gaussian distribution whose parameters are the output of a neural network (\textit{decoder}) with input $z$. 
% The posterior $p(x|z)$ is approximated by $q(z|x)$ which is also a neural network (\textit{encoder}). The two neural networks, decoder and encoder, are trained to maximize a lower bound on the log-likelihood of the data $\log(p(x))$:
% \begin{equation*}
% 	\mathbb{E}_{q(z|x)}[\log p(x|z)]-\textit{KL}(q(z|x) || p(z)),
% \end{equation*}
% where $\textit{KL}$ is the \textit{Kullback-Leibler divergence}.

\subsection{Monte Carlo Attack}\label{subsec:monte_carlo}

In the following section we introduce the first attack which is applicable to all generative models. The intuition behind the Monte Carlo attack is that the generator $G$ overfits if it tends to output datasets close to the provided training data. 
Formally, let $U_\varepsilon(x)$ denote the $\varepsilon$-neighborhood of $x$ defined as $U_\varepsilon(x)=\{x'\,|\,d(x,x')\le \varepsilon\}$ with respect to some distance $d.$ If a sample $g$ of the generative model $G$ is likely to be close to a record $x$ the probability $P(g\in U_\varepsilon(x))$ is increased. It can be rewritten as
%\red{Formally, the probability $P(G\in U_\varepsilon(x))$ is increased where $x$ is a training data record and $G$ is a random variable $G\sim p_{generator}$ denoting the output of the generator. $U_\varepsilon(x)$ denotes the $\varepsilon$ environment of $x$ defined as $U_\varepsilon(x)=\{x'|d(x,x')\le \varepsilon\}$ with respect to some distance $d$. This probability can be rewritten as}
\begin{equation*}
P(g\in U_\varepsilon(x))
=\mathbb{E}_{g\sim p_{generator}} \left(\mathbf{1}_{g\in U_\varepsilon(x)}\right) \label{eqn:prob_eps_neighborhood}
\end{equation*}
and approximated via Monte Carlo integration \cite{mcbook}
\begin{equation}\label{eqn:mc_eps}
\hat{f}_{\mathit{MC}-\varepsilon}(x)=\frac{1}{n}\sum_{i=1}^{n}\mathbf{1}_{g_i\in U_\varepsilon(x)},
\end{equation}
where $g_1,\dots,g_n$ are samples from $p_{generator}$. Note that samples $g_i$ of the generator $G$ are ignored if their distance to the training data record $x$ is higher than $\varepsilon$. In this attack, the estimation $\hat{f}_{\mathit{MC}-\varepsilon}(x)$ plays the role of the function $\hat{f}(x)$ attaining higher values for training data records.

An alternative is provided by incorporating the exact distances $d(z_i,x)$ between samples $g_1,\dots,g_n$ and training data $x$, and computing
\begin{equation*}
\mathbb{E}_{g\sim p_{generator}} \left(-\mathbf{1}_{g\in U_\varepsilon(x)}\log\left(d(g,x) + \delta\right)\right)
\end{equation*}
where a small $\delta$ is chosen to clip off large values ("avoid $\log(0)$") if the distance is zero.
The logarithm is to ensure that outliers do not affect the results too much. The Monte Carlo approximation is then given by
\begin{equation}\label{eqn:mc_dist}
    \hat{f}_{\mathit{MC}-d}(x)
    = -\frac{1}{n}\sum_{i=1}^{n} \mathbf{1}_{g_i\in U_\varepsilon(x)}\log d(g_i,x)  \quad.
\end{equation}
Here, the estimation $\hat{f}_{\mathit{MC}-d}(x)$ plays the role of the function $\hat{f}(x)$ used to conduct the attack types presented above.

In the case of GANs and VAEs one obtains $g_i~\sim~p_{generator}$ by sampling from $z_i\sim p_{noise}$ and computing $g_i=G(z_i)$ and $g_i=D(z_i)$, respectively.
Note that only a sufficiently large amount of samples has to be provided and no additional information is required. 
Of course, both attack variants depend on the specification of the distance $d(\cdot,\cdot)$. See below for details.

%Depending on the type of data used, a meaningful distance metric has to be chosen.
\blue{A further alternative to the attacks discussed could be realized using a Kernel Density Estimator~(KDE)~\cite{parzen1962estimation}.}
In the following we briefly compare the Monte Carlo attack with \blue{this metric.}
%\red{a metric which is frequently used to evaluate generative models \cite{goodfellow2014generative}: the Kernel Density Estimator~(KDE)~\cite{parzen1962estimation}.}
An estimation of the likelihood $\hat{f}(x)$ of a data point $x$ using KDE is given by
\begin{equation}\label{eqn:kde_estimate}
    \hat{f}_{\mathit{KDE}}(x)
    = \frac{1}{nh^d}\sum_{i=1}^{n} K\left(\frac{x-g_i}{h^d}\right),
\end{equation}
where $K$ is typically the Gaussian kernel and $h$ denotes the bandwidth. If this likelihood $\hat{f}_{\mathit{KDE}}(x)$ is significantly higher for training data than for test data the model fails to generalize. Likewise the approximate likelihood values $\hat{f}_{\mathit{KDE}}(x)$ can be used as the function $\hat{f}(x)$ to conduct the single and set MI attack types.
However, this attack variation did not perform better than random guessing and is therefore not considered in our evaluation section.

Note that KDE \eqref{eqn:kde_estimate} can indeed be interpreted as a special case of the proposed distance based method \eqref{eqn:mc_dist}, where
\begin{align*}
    d(x,g_i) &= 1/\exp(h^d\cdot K((x-g_i)/h^d)), \text{ and} \\
    \varepsilon &= \max_{i=1,\dots,n}d(x,g_i) \quad.
\end{align*}
As KDE does not perform well for MI against generative models this stresses that choosing the right distance function seems to be key.
In contrast to KDE, our attacks exclusively consider samples significantly close to training data $x$. 

To fully specify the Monte Carlo attacks concrete distance measures and heuristics for choosing $\varepsilon$ are required. We describe our approach for this in the next two subsections.

\subsubsection{Distance Measures}
\label{subsec:distance_measure}

Both Monte Carlo (MC) attack variants require a distance function $d(\cdot,\cdot)$ and the distance plays an important role for the success of the MI attack.
Therefore, a distance metric suited for the specific data under consideration has to be chosen.
%In our empirical evaluation we used the first two distances for the grayscale images and the first and third for color images. 
%As we will highlight in our evaluation (Section~\ref{sec:experiments}) the
For neural networks, image recognition has become a key task and consequently, we formulate distance metrics for image data in the following paragraphs.

\noindent\textbf{Principal Components Analysis.}
Images are initially represented as a vector of their pixel intensities. A principal component analysis (PCA) is then applied to all vectors in the test dataset. The top 40 components are kept while all other components are discarded. When computing the distance between two new images the PCA transformation is first applied to their vectors of pixel intensities. The Euclidean distance of the two resulting vectors with 40 components each is then defined as the distance of the images.

\noindent\textbf{Histogram of Oriented Gradients.}
Histogram of Oriented Gradients (HOG)~\cite{dalal2005histograms} is a computer vision algorithm enabling the computation of feature vectors for images. First, the image is separated into cells. Second, the occurrences of gradient orientations in the cells are counted and a histogram is computed. The histograms are normalized block-wise and concatenated to obtain a feature vector. Again the Euclidean distance of these vectors is used as image distance. This approach was successfully used by Ebrahimzadeh et al.~\cite{ebrahimzadeh2014efficient} for an MNIST data classifier.

\noindent\textbf{Color Histogram.}
According to the intensities in the three color channels, the pixels are sorted into bins. For the pixels of one image, this results in a color histogram (CHIST) which can be represented as a feature vector. The Euclidean distance of these vectors is defined as the image distance.

\subsubsection{Heuristics for $\varepsilon$} 
For the attack all pairwise distances $d(x_i,g_j)$ of the records $x_i$ and samples $g_j$ need to be computed.  Samples with distances greater than $\varepsilon$ to the training data records are ignored. Hence, an appropriate choice of $\varepsilon$ is crucial for the success of the attack. We thus formulate two heuristics in the following.

\noindent\textbf{Percentile Heuristic.}
The first heuristic is to use a fixed percentile of all pairwise distances $d(x_i,g_j)$ as $\varepsilon$. By choosing the $0.1$\% percentile of the distances as $\varepsilon$ we can ensure that the corresponding samples in an $\varepsilon$-neighborhood are sufficiently close. Note that the MC-$\varepsilon$ and MC-$d$ approaches are not necessarily equivalent if this heuristic is employed.

\noindent\textbf{Median Heuristic.}
The second heuristic avoids the need to choose an additional parameter such as the percentile value. Again, the idea is to exploit the measured distances in the Monte Carlo computation. In this approach, the median of the minimum distance to each record $x_i$ for all the generated samples $g_j$ is chosen:
\begin{equation} \label{eqn:dyn_eps}
    \varepsilon = \median_{1\le i \le 2M}\left(\min_{1\le j\le n} d(x_i,g_j)\right) \quad.
\end{equation}
If $\varepsilon$ is chosen according to the median heuristic \eqref{eqn:dyn_eps} the results of MC-$\varepsilon$ and MC-$d$ are equivalent in both the single and set MI types as there are always exactly $M$ records with $\hat{f}_{\mathit{MC}-\varepsilon}(x_i)>0$ and $\hat{f}_{\mathit{MC}-d}(x_i)>0$.
A comparison of the MC attack variants is provided in the evaluation in Section~\ref{sec:experiments}.

\subsection{Reconstruction Attack}\label{subsec:rec_attack}

The reconstruction attack is solely applicable to VAEs. During training, reconstructions $D(z)$ close to the current training data record $x$ are rewarded. Hence, for training data more precise reconstructions of the VAE can be expected. However, the outputs $D(z)$ are not deterministic. They depend on the latent variable $z$ which is sampled from the distribution $N(E_\mu(x),E_\Sigma(x))$ whose parameters are the output of the encoder network~$E$. Hence, we repeat this process $n$ times and set
\begin{equation}
	\hat{f}_\text{rec}(x)=-\frac{1}{n}\sum_{i=1}^{n} \lVert D(z_i)-x \rVert
\end{equation}
where $z_i$ ($i=1,\ldots, n$) are samples from the distribution $N(E_\mu(x),E_\Sigma(x)).$ This term is frequently used in practice as part of the loss function of VAEs. One of the contributions of this work is to apply this loss to the problem of membership inference. Specifically, the function $\hat{f}_\text{rec}(x)$ is applied in the attack types as the discriminating function $\hat{f}(x)$. This induces the Reconstruction attack. Note that this attack considers a strong adversary $\mathcal{A}$ with access to the VAE model.

% Keine Ahnung, ob wir das sagen sollen: Wie bei Black Box: applicable to other models as well

%\subsection{Mitigation of the Attacks}\label{subsec:mitigation}
%
%The straightforward approach to reduce the vulnerability of (generative) models to MI attacks is to employ standard techniques against over-fitting.
%Possible approaches are data augmentation or merely increasing the size of the training data set. Of course this depends on the availability of additional training data or of the possible ways to augment given data which needs to be adjusted to the type of data the models generate.
%Another known approach is using differential privacy on the training data or during the iterative adjustment of the weights as in \cite{abadi2016dldp}.
%In this paper we provide experimental data about the influence of the training data size and of dropout in Section~\ref{sec:training_data_size}.
%A detailed evaluation of the differential privacy based mitigation of our MI attacks will be done in subsequent work.

% !TeX root = mi_attack_against_gan.tex

\section{Evaluation}
\label{sec:experiments}
The two MI attacks formulated in this paper are evaluated in comparison to the white and black-box MI attacks of Hayes et al.~\cite{Hayes2019} against generative models trained on MNIST, Fashion MNIST, and CIFAR-10 throughout Sections~\ref{sec:mnist} to \ref{sec:cifar}.

%\subsubsection{White- and Black-Box Attack}

The \textit{white box attack} is solely applicable to GANs and requires access to the discriminator $D$. %The intuition behind this attack is that $D$ tends to attain higher outputs $D(x)$ if the record $x$ was part of the training data due to the indirect reward during training. 
Specifically, the discriminator $D$ plays the role of the function $\hat{f}(x)$ in this attack.

%\begin{figure*}[!t]
%	\centering
%	
%	\begin{tikzpicture}[node distance=1cm, auto,]
%	
%	\node[data] (dx) {$D'(x)$};
%	
%	\node[node, left=of dx] (dis) {Discriminator %$D'$}
	%edge[arrow] (dx.west);
	%
	%\node[data, above left=1cm of dis] (fake) %{$G'(z')$}
	%edge[arrow] (dis.west);
	%
	%\node[node, left=of fake] (gen) {Generator $G'$}
	%edge[arrow] (fake.west);
	%
	%\node[data, left=of gen] (z) {$z'\sim p_{noise}$}
	%edge[arrow] (gen.west);
	%
	%\node[data, below left=1cm of dis] (real) %{$g_i\sim p_{generator}$}
	%edge[arrow] (dis.west);
	%
	%\node[node, left=of real] (train) {Target %Generative Model}
	%edge[arrow] (real.west);
	%
	%\node[draw,inner sep=3mm,label=above:Training %Data,fit=(train) (train) (real) (real)] {};
	%
	%\end{tikzpicture}
	%
	%\caption{Schema of the black-box attack.}
%
	%\label{fig:bb_attack}
%\end{figure*}
%
The \textit{black box attack} %\red{, is depicted in Figure~\ref{fig:bb_attack} and} 
overcomes the limitation of the white box attack in that it requires no access to $D$. It is therefore not solely applicable to GANs.
For the black box attack, an auxiliary GAN is trained with samples $g_1,\dots,g_n$ from the target model and the discriminator $D'$ of this newly trained model is used in a white box manner.
%Consequently, records with increased $D'(x)$ are considered part of the training data. Hence, the discriminator $D'$ serves as the function $\hat{f}(x)$ attaining higher values for training than for test data. Though not explicitly tested in the original paper, the black-box attack is also applicable to other generative models such as VAEs since it only requires access to samples $g_1,\dots,g_n$. 
In experiments, the white box attack performed significantly better than the black box attack~\cite{Hayes2019}.

\blue{In general, our MC attacks outperformed state of the art, i.e. the white box attack of Hayes \cite{hayes2017logan}, for both MNIST and Fashion MNIST which are considered very hard datasets due to their simplicity. Since it is an upper bound for the accuracy, also the black box attack is outperformed. However, the MC attacks are dominated by the white box attacks on CIFAR-10. This is due to the bad sample quality which is essential if only very close samples are considered. As a consequence of the low accuracies, we decided not to compare it with the black-box attacks. In contrast, the Reconstruction attack specialized for VAEs constantly provides the highest accuracies with up to 100\% single and set accuracies even for CIFAR-10.}

Since several parameters have to be chosen before the attacks are applied a study of the effect of these parameters is presented in Section~\ref{sec:params}.
Moreover, additional experiments on VAEs trained on the MNIST dataset are provided in Sections~\ref{sec:subsets} and \ref{sec:training_data_size}. These experiments are not performed for the other datasets or GANs to avoid redundancy and are solely for the purpose of evaluating the effect of regularization and training data sizes.

\subsection{Setup}
\label{subsec:setup}

We evaluated the attacks of Hayes et al.~\cite{hayes2017logan}, the Monte Carlo and the Reconstruction attacks for differing 10\% subsets of the MNIST, Fashion MNIST and the CIFAR-10 dataset. 
While the simple nature of MNIST has proven to result in low MI precision in previous work, the more complex Fashion-MNIST and CIFAR-10 datasets result in higher MI precision. Thus, the three chosen datasets represent three varying difficulties w.r.t.~MI.
%The remaining 90\% are used as test set in both the single and set MI attack type.
To ensure a fair comparison we executed all experiments repeatedly and report standard deviations.
Neural networks are implemented with tensorflow \cite{abadi2016tf}, and for the HOG and PCA computations, the python libraries scikit-image and scikit-learn \cite{scikit-learn} are used. Experiments were run on Amazon Web Services p2.xlarge (GAN) and c5.2xlarge (VAE) instances.

We first describe the datasets and models used before analyzing the parameters of the attacks.

\subsubsection{MNIST}

MNIST is a standard dataset in machine learning and computer vision consisting of $70,000$ labeled handwritten digits which are separated into $60,000$ training and $10,000$ test records.\footnote{http://yann.lecun.com/exdb/mnist/} Each digit is a $28\times28$ grayscale image. In all subsequent datasets only a $10\%$ subset of the training images is used for training to provoke overfitting. The remaining $90\%$ of the training data is used as test data to compute the accuracies of the attacks. The actual MNIST test data is only used to define the PCA transformation for the PCA based distance. This ensures that the distance is not influenced by the specific choice of the training data or the remaining $90\%.$
Attacks are performed against two state of the art generative models, namely GANs~(cf. Section~\ref{sec:gan}) and VAEs~(cf. Section~\ref{sec:vae}). For the GAN we employ the widely used deep convolutional generative adversarial network (DCGAN)~\cite{radford2015unsupervisedRL} architecture which aims to improve both stability and quality of GANs for image generation. This network relies on convolutional neural networks (CNN) which are state of the art for many computer vision tasks.
We trained the DCGAN for $500$ epochs (i.e., until convergence) with a mini batch size of $128$.\footnote{We used https://github.com/yihui-he/GAN-MNIST as a starting point.}
For the VAE we apply a standard architecture\footnote{We used https://github.com/hwalsuklee/tensorflow-mnist-VAE as a starting point.} with $90\%$ Dropout and a mini batch size of $128$. Due to the different convergence behavior, the VAE is only trained for $300$ epochs. For both models, GAN and VAE, we utilize the conditional variant s.t. we can control which digit is generated.

\subsubsection{Fashion MNIST} This dataset is intended to serve as a direct drop-in replacement for MNIST~\cite{xiao2017fashion}. Like MNIST it consists of $60,000$ training and $10,000$ test $28 \times 28$ grayscale images representing $10$ fashion classes such as trousers, pullovers etc. The goal of using this dataset is to overcome the limitation of MNIST being too simple for various computer vision tasks. The same model architectures as that for MNIST are used for the conditional GAN and VAE on this dataset.

\subsubsection{CIFAR-10}

The CIFAR-10 dataset~\cite{krizhevsky2009learning} consists of $60,000$ $32\times32$ color images representing $10$ classes such as airplane, automobile etc. There are $50,000$ train and $10,000$ test records. 
Within the evaluation a GAN\footnote{We used https://github.com/4thgen/DCGAN-CIFAR10 as a starting point.} and a VAE\footnote{We used https://github.com/chaitanya100100/VAE-for-Image-Generation as a starting point.} are trained on a random $10\%$ subset of the original dataset.

\subsection{Attack Parameters}\label{sec:params}

The effects of the attack parameters are analyzed in the following. Specifically, for the MC attacks the effect of the heuristic for setting $\varepsilon$ and the number of samples $n$ for the Monte Carlo integration are studied. We expect these to be similar for both GANs and VAEs. Hence, the analysis is restricted to the case of VAEs. 
For the Reconstruction attack, we study how the number of samples $n$ for the reconstruction error estimation affects the accuracy.

\subsubsection{Monte Carlo Attack}

The single and set MI accuracies  against VAEs trained on MNIST for different choices of $\varepsilon$ are reported in Table~\ref{tab:eps} for \blue{$\mathcal{A}$} and \blue{$\mathcal{R}$}, respectively. Note that the results of the MC-$\varepsilon$ and MC-$d$ attacks do not differ significantly. This suggests that the main contribution is the introduction of $\varepsilon$ effectively ignoring samples which are further than $\varepsilon$ away from the training records. In the case of the median heuristic, the two MC attack variants yield equivalent performances as expected. However, the median heuristic outperforms the percentile heuristic.

\begin{table*}[!htb]
	\centering
	\caption{Set accuracies for \blue{$\mathcal{R}$} depending on $\varepsilon$ values}
	\label{tab:eps}
	\subfloat[HOG-based distance]{
		\centering
		\begin{tabular}{ccccc}
			\hline
			Heuristic/Percentile & \multicolumn{4}{c}{HOG-based distance} \\
			& GAN Monte Carlo-d & GAN Monte Carlo-$\varepsilon$& VAE Monte Carlo-d & VAE Monte Carlo-$\varepsilon$\\
			\hline
			Median &  63.76$\pm$3.83 & 63.76$\pm$3.83 & 83.50$\pm$2.43 & 83.50$\pm$2.43\\
			0.01\% & 63.76$\pm$3.68 & 66.11$\pm$3.70 & 81.00$\pm$2.59 & 82.25$\pm$2.50\\
			0.10\% & 63.76$\pm$3.71 & 62.08$\pm$3.65 & 74.50$\pm$2.90 & 71.75$\pm$2.98\\
			1.00\% & 60.07$\pm$3.84 & 59.73$\pm$3.86 & 59.50$\pm$3.24 & 54.00$\pm$3.29\\
			\hline
		\end{tabular}
	}%
	\\
	\subfloat[PCA-based distance]{
		\centering
		\begin{tabular}{ccccc}
			\hline 
			Heuristic/Percentile & \multicolumn{4}{c}{PCA-based distance} \\
			& GAN Monte Carlo-d & GAN Monte Carlo-$\varepsilon$& VAE Monte Carlo-d & VAE Monte Carlo-$\varepsilon$\\
			\hline
			Median &   74.84$\pm$3.25 & 74.84$\pm$3.25 & 99.75$\pm$0.25 & 99.75$\pm$0.25\\
			0.01\% &  74.84$\pm$3.31 & 71.94$\pm$3.40 & 95.50$\pm$1.34 & 91.75$\pm$1.80\\
			0.10\% &  64.84$\pm$3.69 & 59.68$\pm$3.78 & 94.75$\pm$1.52 & 95.50$\pm$1.43\\
			1.00\% &  47.42$\pm$3.77 & 51.61$\pm$3.76 & 60.75$\pm$3.21 & 58.50$\pm$3.29\\
			\hline
		\end{tabular}
	}%
\end{table*}

\begin{figure*}[!t]
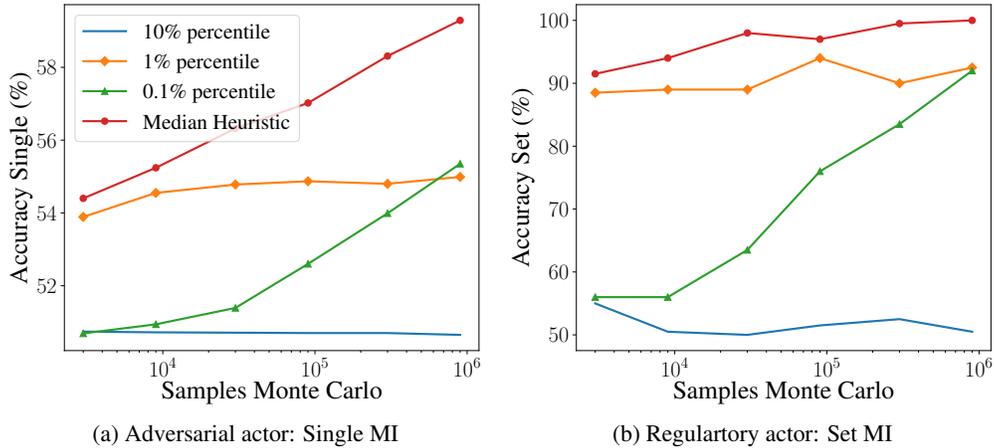

	\centering
	\subfloat[Adversarial actor: Single MI]
	{
		\scalebox{0.4}{\input{img/Monte_Carlo_Samples.pgf}}
	}
	\subfloat[Regulartory actor: Set MI]
	{
		\scalebox{0.4}{\input{img/Monte_Carlo_Samples_Set.pgf}}
	}
	\caption{MC attack accuracy (differing scales) on MNIST with PCA based distance against VAEs depending on sample size.}
	\label{fig:acc_sample_size}
\end{figure*}

Besides the heuristic for $\varepsilon,$ a sample size for the Monte Carlo approximation has to be chosen. Hence, we also analyze the performance of the MC-$\varepsilon$ attack depending on the sample size. Again, the MC-$\varepsilon$ attack is equivalent to the MC-$d$ attack in the case of the median heuristic. The single and set accuracies are stated in Figure~\ref{fig:acc_sample_size} for \blue{$\mathcal{A}$} and \blue{$\mathcal{R}$}, respectively.
In general, higher percentile values ignore fewer samples since $\varepsilon$ is increased. A smaller sample size is required to achieve optimal accuracy for these percentiles. However, the accuracy of higher percentile values is inferior to the ones of lower percentile values.

For example, the $10\%$ percentile attack already reaches its optimum in the minimal case of $3,000$ samples and the $1\%$ percentile saturates at $10^4$ samples. The $0.1\%$ percentile approach is gaining higher accuracies and does not level off at $10^6$ samples. It is noticeable that the median heuristic always outperforms the other heuristics. We conjecture this heuristic to level off at a higher sample size. However, in practice there is a trade-off between computational effort and accuracy of the attack. To study the effect $20$ experiments for the median heuristic with $10^7$ samples each are conducted, achieving a single record MI accuracy of $59.80\pm 3.50\%$ for \blue{$\mathcal{A}$} and a set MI accuracy of $100.00 \pm 0.00\%$ for \blue{$\mathcal{R}$}. In the subsequent experiments, we always use $10^6$ samples for the Monte Carlo simulations.

The median heuristic is superior to the percentile heuristic for all sample sizes. Moreover, no parameter like the percentile is required. Thus, in all subsequent experiments we apply the median heuristic for which the MC-$\varepsilon$ and MC-$d$ attacks are equivalent. We refer to these equivalent approaches simply as \textit{MC attack}.

\subsubsection{Reconstruction Attack} We also study the effect of the sample size $n$ to approximate the reconstruction error 
\begin{equation}
\hat{f}_\text{rec}(x)=-\frac{1}{n}\sum_{i=1}^{n} \lVert D(z_i)-x \rVert.
\end{equation}
In preliminary experiments even small sample sizes of $n=300$ yielded good accuracies. This suggests that the estimator $\hat{f}_\text{rec}(x)$ is accurate enough for small $n$ values. To ensure optimal results we conduct the subsequent experiments with $n=10^6$ for the Reconstruction attacks against VAEs trained on MNIST and Fashion MNIST. For CIFAR we just use $n=10^5$ samples as we already achieve accuracies of $\approx100\%$ both in single and set MI.

\subsection{Results on MNIST}
\label{sec:mnist}

Having analyzed the parameters of our proposed attacks, we now compare their accuracies with the recent white-box and black-box attacks of~\cite{Hayes2019}. 
To stabilize the results $10$ different $10\%$ subsets of the MNIST data are chosen as training data for the GAN and VAE models. For every subset $10$ single and set MI attacks are conducted with $M=100$. While we apply the white-box attack against the GAN, we are limited to the black-box attack in case of the VAE as the latter model does not feature a discriminator. In order to test the black-box attack, a new GAN is trained with $10^6$ samples from the target VAE.

For the Monte Carlo estimator $\hat{f}_{\mathit{MC}}$ we use the PCA and HOG based distances introduced in Section~\ref{subsec:distance_measure}. The CHIST distance is not applicable since MNIST solely consists of grayscale images. As described in the previous section we use $n=10^6$ samples and the median heuristic. The resulting accuracies are depicted in Figure~\ref{fig:results_vae_gan}. The dotted horizontal baseline at $50\%$ is the average success rate of random guessing.
In general, the accuracies of single MI for \blue{$\mathcal{A}$} are significantly lower than those of set MI for \blue{$\mathcal{R}$}. Furthermore, all attacks are much more successful if applied against VAEs instead of GANs. This suggests that in general there is less overfitting in GANs. This observation is consistent with the Annealed Importance Sampling measurements by Wu et al.~\cite{wu2016quantitative}. 

The black-box and white-box attack do not perform significantly better than the baseline in both experiments. The MC attack clearly outperforms these attacks in the experiments. When used with PCA distance our MC attack can even infer set membership with nearly $100\%$ accuracy against a VAE. 
For the GAN the accuracy is still about $75\%$. In general, accuracies are inferior if the HOG distance is used.
As a side fact, the Monte Carlo based attacks with PCA distance take $\approx 7$ minutes each on a p2.xlarge instance on AWS. Currently, at the cost of $0.90$ US \$ per hour, the attacks only cause minor costs.
The specialized Reconstruction attack is superior to the MC attack in the case of the VAE yielding $\approx70\%$ and $100\%$ in the single and set MI attack, respectively. 
The high accuracies of the attacks we proposed make them especially attractive for the regulatory use case depicted in Section~\ref{sec:attack_scenarios}.

% Updated for dynamic eps
%\begin{figure*}[!t]
%	\centering
%	\subfloat[Single Membership Inference]
%	{
%		\scalebox{0.4}{\input{img/GAN_VAE_Single.pgf}}
%	}
%	\subfloat[Set Membership Inference]
%	{
%		\scalebox{0.40}{\input{img/GAN_VAE_Set.pgf}}
%	}
%	\caption{Average accuracy (differing scales) of the attacks on MNIST in the single and set experiments with standard deviation.}
%	\label{fig:results_vae_gan}
%\end{figure*}

% Note that the MC attack performs much better if the PCA based distance is used. Therefore, it is also crucial which distance is chosen. The distance metric has to be chosen according to the data of a specific problem (e.g., image data vs. audio data). Observe that the two main changes between the naive Euclidean approach and the proposed MC attack are that different distances are used and only the very small distances are considered depending on the specific choice of $\varepsilon$. This suggests that overfitting in generative models can be identified by looking at very close samples with respect to a suitable distance metric. 

\subsection{Effect of Subset Choice}\label{sec:subsets}

\begin{figure*}[!t]
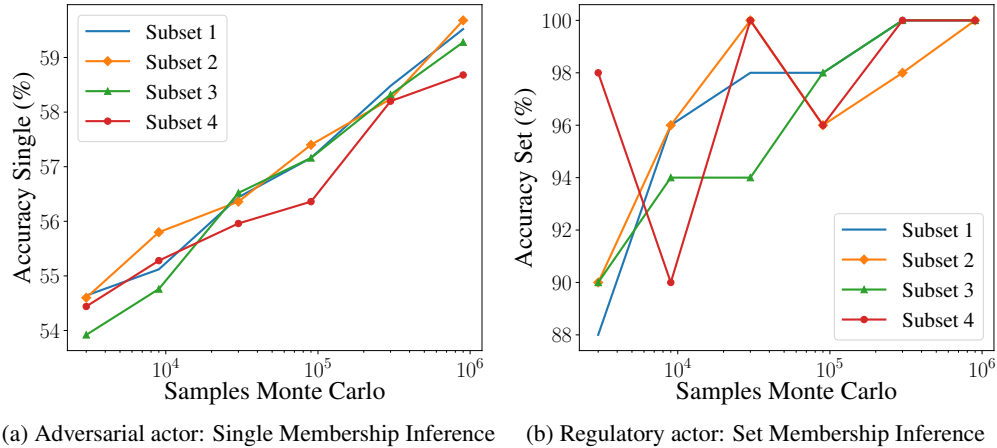

	\centering
	\subfloat[\blue{Adversarial actor:} Single Membership Inference]
	{
		\scalebox{0.4}{\input{img/Monte_Carlo_Samples_Subset.pgf}}
	}
	\subfloat[\blue{Regulatory actor:} Set Membership Inference]
	{
		\scalebox{0.40}{\input{img/Monte_Carlo_Samples_Set_Subset.pgf}}
	}
	\caption{MC attack accuracy (differing scales) on MNIST with PCA distance depending on sample size for four different training subsets.}
	\label{fig:subset_effect}
\end{figure*}

It is unclear how the specific choice of the MNIST $10\%$ subset influences the accuracy of the MC attack. In Figure~\ref{fig:subset_effect} the average MC attack performance with PCA distance against VAEs trained on different subsets are plotted. 
Attack performances seem independent of the specific subset. We also conduct an $F$-test to evaluate whether the single accuracy means of the four VAEs are different at $10^6$ samples resulting in a $p\text{-value}\approx0.64$. 
Hence, the hypothesis that the means are equal can be accepted with high probability, i.e.\ the choice of the subset does not significantly influence the attack results. We conclude that the accuracy depends on the size of the training data rather than its specific members.

We remark that in the experiment setups $M=100$ samples of the $10\%$ subset of the training data and $100$ samples of the remaining $90\%$ training data are chosen. The set MI experiments yield high accuracies.
Therefore, if a regulator suspects that some dataset was used for training a model this can be recognized with the novel attacks even though other data might have been part of the training data as well. This is an analogous case to the experiment described. Though of course more training data was used, we focus on $100$ samples. It is very likely that the inappropriately used data is not the only data used to train the model. Hence, the practicability of the MC attack is increased since the regulator does not need to know all the training data to prove that a certain subset was used.

\subsection{Effect of Training Data Size and Regularization --- Mitigations}\label{sec:training_data_size}

We also investigate how the size of the training dataset influences the success of the attacks for the MNIST dataset. For this, five VAEs are trained with $20$ experiments each since the effect should be similar for GANs. The results for the MC attack and Reconstruction attack are depicted in Table~\ref{tab:tds}. 
When using $40\%$ of the training data instead of the usual $10\%$ the accuracy shrinks from $60\%$ to $51\%$ for single MI and from nearly $100\%$ to only about $58\%$ for set MI in the case of the MC attack. As expected, for $20\%$ the effects are less significant. Clearly,  more training data would further reduce the effectiveness of the attacks. However, in the case of the Reconstruction attack, the effects are less significant. Even if $40\%$ are used the set accuracy is still about $100\%$ meaning that the Reconstruction attack is more robust.

In general, the performance declines with more training data suggest that generative models make use of the additional information provided by additional training data. Similar effects were observed before in the case of the white-box attack~\cite{Hayes2019}.
\begin{table}
	\centering
	\caption{Accuracies depending on MNIST training data size}
	\label{tab:tds}
	\begin{tabular}{cc@{\hspace{0.5em}}cc@{\hspace{0.5em}}c}
		\hline
		Size & \multicolumn{2}{c}{Monte Carlo (PCA dist.)} & \multicolumn{2}{c}{Reconstruction attack}\\ 
		 & Single & Set & Single & Set\\
		\hline
		40\% & 50.79$\pm$0.27 & 57.50$\pm$3.24 & 57.35$\pm$0.37 & 98.50$\pm$1.11\\
		20\% & 57.05$\pm$0.32 & 94.75$\pm$1.39 & 62.23$\pm$0.38 & 100.00$\pm$0.00\\
		10\% & 59.93$\pm$0.26 & 99.75$\pm$0.25 & 70.09$\pm$0.37 & 100.00$\pm$0.00\\
		\hline
	\end{tabular}
\end{table}
However, often in practice the amount of training data is a bottleneck for training generative models. In consequence, one could use regularization methods to improve the generalization such as \textit{dropout}~\cite{srivastava2014dropout}. In the case of dropout, certain neurons are switched off during training with given probability to increase the resistance of the network. In the standard case we already use dropout with a keep probability of $90\%$ both in the encoder and decoder of the VAE. We also conduct experiments for the MC and Reconstruction attack at lower keep rates of $70\%$ and $50\%$. The accuracy in the set MI type decreases to $79\%$ at a keep probability of $70\%$ and to $65\%$ at an even reduced keep probability of $50\%$ for the MC attack. Again, the effects are less significant for the Reconstruction attack still yielding $\approx86\%$ set MI accuracy for a $50\%$ keep rate. Detailed results are reported in Table~\ref{tab:dropout}. 
The results indicate that dropout can indeed be used in practice to mitigate the proposed MI attacks. This can also be observed in the case of the white-box attack~\cite{Hayes2019}. However, a lower keep probability also causes the generated images to get increasingly blurry (cf.~Appendix, Figure~\ref{fig:generated_img_vae_gan_mnist}). Hence, there is an inherent trade-off between high image quality and low MI attack accuracies.

\begin{table}
	\centering
	\caption{Accuracies depending on MNIST Dropout Keep Rates}
	\label{tab:dropout}
	\begin{tabular}{cc@{\hspace{0.5em}}cc@{\hspace{0.5em}}c}
		\hline
		Rate & \multicolumn{2}{c}{Monte Carlo (PCA dist.)} & \multicolumn{2}{c}{Reconstruction attack}\\ 
		& Single & Set & Single & Set\\
		\hline
		50\% & 51.45$\pm$0.26 & 64.75$\pm$3.19 & 53.77$\pm$0.34 & 86.00$\pm$3.18\\
		70\% & 53.17$\pm$0.29 & 78.50$\pm$2.71 & 58.31$\pm$0.40 & 97.00$\pm$1.56\\
		90\% & 59.93$\pm$0.26 & 99.75$\pm$0.25 & 70.09$\pm$0.37 & 100.00$\pm$0.00\\
		\hline
	\end{tabular}
\end{table}

\subsection{Results on Fashion MNIST}

Samples of the trained VAE and GAN models are provided in Figure~\ref{fig:generated_img_vae_gan_fashion} (Appendix). They show that the GAN produces more detailed samples compared to the VAE. 

\begin{figure*}[!t]
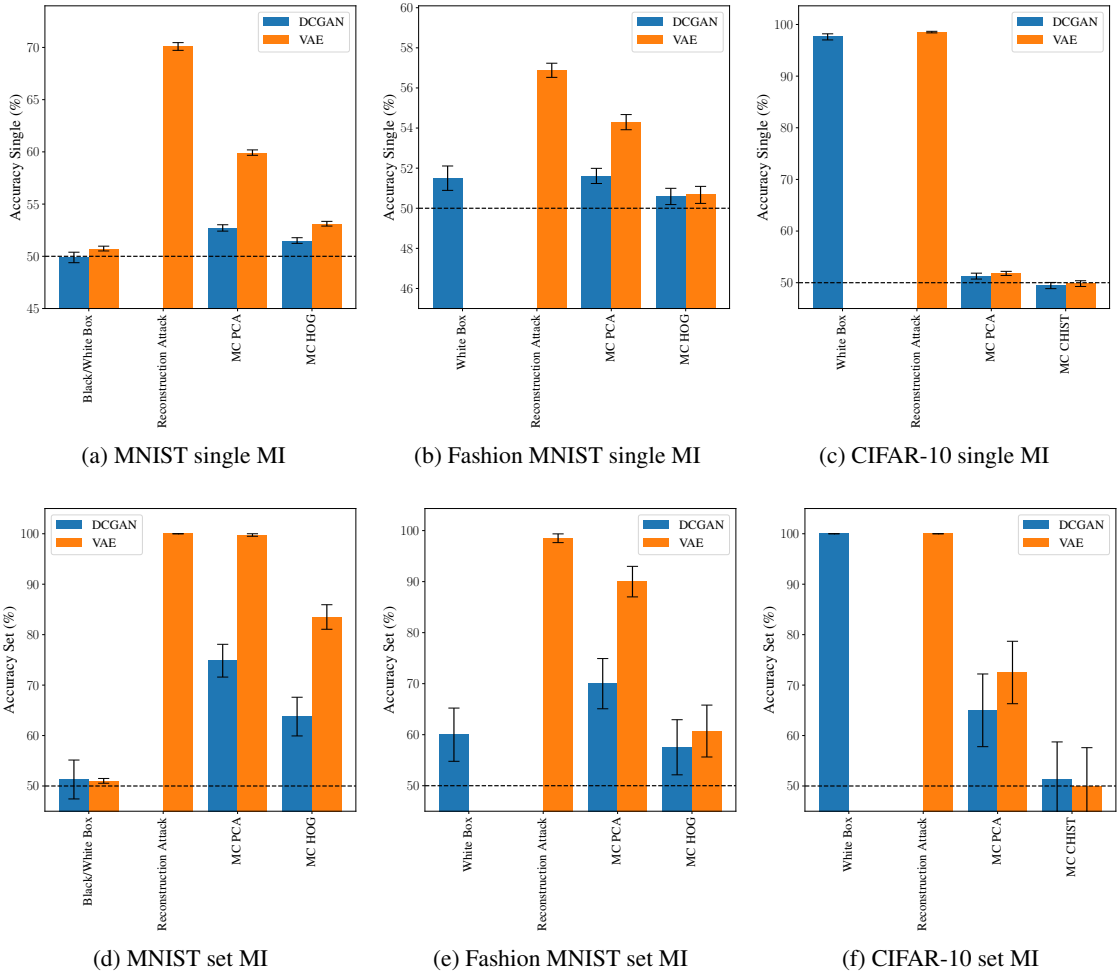

	\centering
		\subfloat[MNIST single MI]
		{
			\scalebox{0.3}{\input{img/GAN_VAE_Single.pgf}}
		}
		\subfloat[Fashion MNIST single MI]
		{
			\scalebox{0.3}{\input{img/Fashion_GAN_VAE_Single.pgf}}
		}
		\subfloat[CIFAR-10 single MI]
		{
			\scalebox{0.3}{\input{img/CIFAR_GAN_VAE_Single.pgf}}
		}
	\vfill	
		\subfloat[MNIST set MI]
		{
			\scalebox{0.3}{\input{img/GAN_VAE_Set.pgf}}
		}
		\subfloat[Fashion MNIST set MI]
		{
			\scalebox{0.3}{\input{img/Fashion_GAN_VAE_Set.pgf}}
		}
		\subfloat[CIFAR-10 set MI]
		{
			\scalebox{0.3}{\input{img/CIFAR_GAN_VAE_Set.pgf}}
		}
	\caption{Average attack accuracy (differing scales) for single and set MI on the datasets.\\}
	\label{fig:results_vae_gan}
\end{figure*}

To stabilize our results we train five GANs and VAEs on different $10\%$ subsets of the dataset. For each model $20$ single record MI and set MI experiments are conducted. We do not evaluate the black-box attack for the VAE as it performed significantly worse than the MC attack and Reconstruction attack in the previous MNIST experiments. The white-box attack is not applicable since VAEs do not provide a discriminator $D$. Figure~\ref{fig:results_vae_gan} provides an overview of the results.

%\begin{figure*}[!t]
%	\centering
%	\subfloat[Single Membership Inference]
%	{
%		\scalebox{0.4}{\input{img/Fashion_GAN_VAE_Single.pgf}}
%	}
%	\subfloat[Set Membership Inference]
%	{
%		\scalebox{0.40}{\input{img/Fashion_GAN_VAE_Set.pgf}}
%	}
%	\caption{Average attack accuracy (differing scales) on Fashion MNIST with standard deviation.\\}
%	\label{fig:fashion_results_vae_gan}
%\end{figure*}

Compared to MNIST, the MC attack performs slightly worse on this dataset. As before, the attacks are more successful in against the VAE providing additional evidence that GANs generalize better. This surprises because the samples created by the GAN are more detailed. 
The white-box attack performs better with this dataset achieving about $60\%$ accuracy for set MI against GANs. However, it is still inferior to the proposed MC attacks with PCA distance ($70\%$ accuracy).
Again, our reconstruction attack significantly outperforms all other attacks in the case of the VAE yielding $\approx57\%$ and $\approx99\%$ in the single and set case.

\begin{table*}[!htb]
\centering
    \caption{Accuracy of the white-box, Reconstruction and MC Attacks on Fashion MNIST for single record MI and set MI.}
   	\label{tab:fashion}

    \subfloat[Single MI]{
      \centering
	\begin{tabular}{ccccc}
	\hline
		Model & \multicolumn{4}{c}{Accuracy Single (\%)} \\
		 & White-box attack & Reconstruction attack & Monte Carlo PCA distance & Monte Carlo HOG distance\\
		\hline
		GAN & 51.50$\pm$0.61 & not applicable & 51.61$\pm$0.38 & 50.59$\pm$0.41 \\
		VAE & not applicable & 56.88$\pm$0.35 & 54.29$\pm$0.38 & 50.67$\pm$0.43 \\
		\hline
	\end{tabular}
    }
	\\
	 \subfloat[Set MI]{
      \centering
	\begin{tabular}{ccccc}
				\hline
		Model & \multicolumn{4}{c}{Accuracy Set (\%)} \\
		& White-box attack & Reconstruction attack & Monte Carlo PCA distance & Monte Carlo HOG distance\\
		\hline
		GAN & 60.00$\pm$5.22 & not applicable & 70.00$\pm$4.92 & 57.53$\pm$5.40\\
		VAE & not applicable & 98.50$\pm$0.86 & 90.00$\pm$2.99 & 60.71$\pm$5.09 \\
		\hline
	\end{tabular}
    }%
\end{table*}
%\begin{table*}
%	\centering
%	\caption{Accuracy of the White box, AIS and MC Attacks on Fashion MNIST for single record MI and set MI.}
%	\label{tab:fashion}
%	\begin{tabular}{ccccc}
%		\hline
%		Model & \multicolumn{4}{c}{Accuracy Single (\%)} \\
%		 & White Box & AIS & MC PCA & MC HOG\\
%		\hline
%		GAN & 51.50$\pm$0.61 & 50.39$\pm$0.56 & 51.61$\pm$0.38 & 50.59$\pm$0.41 \\
%		VAE & not applicable & 50.13$\pm$0.60 & 54.29$\pm$0.38 & 50.67$\pm$0.43 \\
%		\hline
%		Model & \multicolumn{4}{c}{Accuracy Set (\%)} \\
%		& White Box & AIS & MC PCA & MC HOG\\
%		\hline
%		GAN & 60.00$\pm$5.22 & 53.21$\pm$5.11 & 70.00$\pm$4.92 & 57.53$\pm$5.40\\
%		VAE & not applicable & 51.28$\pm$5.40 & 90.00$\pm$2.99 & 60.71$\pm$5.09 \\
%		\hline
%	\end{tabular}
%\end{table*}

\subsection{Results on CIFAR-10}
\label{sec:cifar}

Samples of the models after training are provided in Figure~\ref{fig:generated_cifar_vae_gan}. Though state of the art models are applied, they do not succeed in learning the data effectively as the samples are very blurry and real objects cannot be identified. This is similar to Hayes et al.~\cite{Hayes2019}. Hence we expect the MC attacks to perform worse on these datasets due to their reliance on samples which are very close to the training data. However, when the overall quality is bad we do not expect individual samples to replicate the training data.

% ###
% Figure (Generated images of a GAN and a VAE after training on CIFAR-10 dataset) shifted to the appendix
% ###

\begin{table*}[t]
	\centering
	\caption{CIFAR-10 accuracy for MC and White-box attack}
	\label{tab:cifar}
	\subfloat[Accuracy of single MI]{
	\begin{tabular}{l@{\hspace*{2em}}cccc}
		\hline
		Model & \multicolumn{4}{c}{Accuracy Single (\%)} \\
		 & White-box attack & Reconstruction attack & Monte Carlo PCA  distance& Monte Carlo CHIST distance\\
		\hline
		GAN & $97.60\pm0.59$ & not applicable & $51.28\pm0.57$ & $49.45\pm0.60$ \\
		VAE & not applicable & $98.52\pm0.15$ & $51.80\pm0.40$ & $49.83\pm0.55$ \\
		\hline
	\end{tabular}
	}
	\\
	\subfloat[Accuracy of set MI]{
	\begin{tabular}{l@{\hspace*{2em}}cccc}
		\hline
		Model & \multicolumn{4}{c}{Accuracy Set (\%)} \\
		& White-box attack & Reconstruction attack & Monte Carlo PCA  distance& Monte Carlo CHIST distance\\
		\hline 
		GAN & $100.00\pm0.00$ & not applicable & $65.00\pm7.21$ & $51.25\pm7.49$\\
		VAE & not applicable & $100.00\pm0.00$ & $72.50\pm6.19$ & $50.00\pm7.60$ \\
		\hline
	\end{tabular}	
	}
\end{table*}
%
%
%
%\begin{table*}[t]
%	\centering
%	\caption{CIFAR-10 accuracy for MC and White Box Attack}
%	\label{tab:cifar}
%	\begin{tabular}{ccccc}
%		\hline
%		Model & \multicolumn{4}{c}{Accuracy Single (\%)} \\
%		 & White Box & AIS & MC PCA & MC CHIST\\
%		\hline
%		GAN & $97.60\pm0.59$ & not evaluated & $51.28\pm0.57$ & $49.45\pm0.60$ \\
%		VAE & not applicable & not evaluated & not evaluated & not evaluated \\
%		\hline
%		Model & \multicolumn{4}{c}{Accuracy Set (\%)} \\
%		& White Box & AIS & MC PCA & MC CHIST\\
%		\hline
%		    
%		GAN & $100.00\pm0.00$ & not evaluated & $65.00\pm7.21$ & $51.25\pm7.49$\\
%		VAE & not applicable & not evaluated & not evaluated & not evaluated \\
%		\hline
%	\end{tabular}
%\end{table*}
MC distances are calculated by the known PCA based distance with $120$ components. Moreover, we examine the CHIST distance (cf. Section~\ref{subsec:distance_measure}) instead of the HOG distance for two reasons. First, the images are very blurry so it is very unlikely that oriented gradients yield a good distance. Second, it is now possible to employ the CHIST distance as it relies on colors and could potentially be less affected by blurry images.

Contrary to the $100$ experiments for MNIST and Fashion MNIST, $40$ experiments were sufficient for significant results for CIFAR-10. The results of the white-box attack and the novel MC and Reconstruction attacks are depicted in Table~\ref{tab:cifar}. Figure~\ref{fig:results_vae_gan} provides an overview of the results. The MC attack with CHIST distance is not significantly better than random guessing. If the PCA based distance is employed the accuracy increases to roughly $51\%$ and $52\%$ for single MI and $65\%$ and $73\%$ set MI against the GAN and VAE, respectively. 
Again, the choice of the distance metric $d$ is crucial. Surprisingly the attack exhibits an accuracy better than random guessing despite the bad sample quality. However, unlike the MNIST and Fashion MNIST datasets, the white-box attack outperforms the MC attack for the GAN trained on CIFAR-10. This is most likely due to the bad sample quality of the generator. 

The white-box attack achieves an accuracy of nearly $100\%$ in single record MI as well as set MI implicating that despite the bad sample quality the discriminator effectively remembers the training data. A similar accuracy can be observed for the reconstruction attack in the case of the VAE.
This suggests that the reconstruction attack we propose is an effective means of assessing VAEs as it constantly outperformed all other attacks. Note that for GANs the white-box attack cannot play this role as it performs worse than the novel MC attacks on MNIST and Fashion MNIST.

\section{Related Work}
\label{sec:related}
% (jetzt als eigenes (vorletztes) Kapitel)
The range of attacks against neural networks and their applications is wide and various approaches have been contributed. We now review the prior work and relate it to our findings.

In the case of adversarial examples, input data is systematically manipulated to disturb inference as formulated by Huang et al.~\cite{Huang2011AdversarialML}. In the case of adversarial training, sample data is poisoned, e.g., to introduce stealthy features which may be exploited later on \cite{mozaffari2015systematic, yang2017generative}.
Common to these examples is an attacker who actively influences the result of either learning or inference of a model.

In contrast, this work considers an honest-but-curious adversary having access to an already trained model, or at least to samples from a generative model. This adversary infers knowledge about the training data records.
Previous work in this setup follows two main directions:
Model inversion attacks as formulated by Fredrikson et al.~\cite{fredrikson2015model} and Tramer et al.~\cite{tramer2016stealing} try to directly reconstruct training data based on the output of a model to which the attacker has black-box access. Instances of this approach can make use of a confidence score for the output in a discriminative model \cite{fredrikson2015model}.

% shokri oder hayes related work
Our approach follows the other main direction of data leakage attacks: membership inference. The goal of this attack is to identify the data used to train the model. Shokri et al.~\cite{shokri2017membership} apply such attacks against discriminative networks. We focus on generative models similar to Hayes et al.~\cite{Hayes2019}, and also evaluate our attacks in comparison to their white- and black-box attacks. The white-box attack, where the discriminator of the trained model must be accessible, is restricted to GANs. The black box attack solely requires access to samples from the model.
We further structure the class of membership inference attacks by assuming \blue{two different types of actors: an honest-but-curious adversary $\mathcal{A}$ performing single MI, and a regulatory actor $\mathcal{R}$ performing set MI}. 
The first attack type has already been used in previous work to evaluate attacks against generative models~\cite{Hayes2019}.
%
% vvvvv changes (MH) related to "2. Novelty of Set MI" vvvvv
%\red{The set MI attack \blue{performed by $\mathcal{R}$} has also been approached by Liu et al.~\cite{LLG18} as co-membership attack of size $n$. However, while we amplify subtle differences in the discriminating function $\hat{f}$, their approach trains a neural network to find possible pre-images for $n$ given outputs of the GAN simultaneously.}
\blue{In parallel to our work, Liu et al.~\cite{LLG18} came up with an approach for the application of MI to a set of samples simultaneously.
Their approach is to train a network A that acts as an inverse for the generator and they then measure the (L2-)distance of the generator applied to the thus calculated preimage of a sample to the sample itself.
The decision to classify a sample as training data is based on a threshold applied to this distance. In their co-membership inference attack they simultaneously train and evaluate the network A on multiple samples (either all training data or all test data). Hence their decision function implicitly changes for different input data.
However, our set membership inference provides a framework where a discriminating function $f$ (which is fixed per attack) is evaluated by $\mathcal{R}$ on a the members of two sets of samples (from training resp. test data) in order to amplify subtle differences in the values of $f$ and to compensate for outliers.
}
% AAAAA ---------------------------------------------AAAAA
%
%From Wu et al.~\cite{wu2016quantitative} we owe the idea of an effective estimation of log-likelihoods by means of annealed importance sampling. They use this for estimating the degree of overfitting of their models. A similar approach was already used by Salkhutdinov et al.~\cite{salakhutdinov2008quantitative} for deep belief networks. \todo{renmove}

% Theis 2016 overfitting (why not use KDE), trifft das auch auf unseren Ansatz zu?
Part of our work can be seen as a generalization of previous approaches to evaluate generative models. According to Theis et al.~\cite{Theis2016a} the choice of metrics may have a strong influence on the result of such model evaluations. Specifically, the use of KDE is problematic since the error may be large. Hence, Theis et al.~\cite{Theis2016a} suggest not to use KDE for the evaluation of generative models. A key difference of our MC attack in comparison to KDE is that it only considers samples very close to the training data. Arora et al. \cite{arora2017generalization} recently evaluated GANs by analyzing near duplicate samples of GANs with the Birthday paradox. Their results lead to the similar conclusion that close samples are of high interest to assess the model quality.

Model quality is related to overfitting. Yeom et al.~\cite{yeom2017unintended} study the relationship between overfitting and the success of both membership inference and model inversion attacks and quantify the advantage of them. Opposed to our work, their analysis considers discriminative models.
We could empirically show a similar effect for generative models. Overfitting increased the accuracy of all examined attacks. This aligns with the results of Hayes et al.~\cite{Hayes2019} for their white-box attack.
% empirisch 

We use histograms of oriented gradients (HOG)~\cite{dalal2005histograms}, color histograms and PCA to quantify distances between images. A different approach would be an algorithm built upon local key point descriptors such as the scale-invariant feature transform (SIFT) algorithm~\cite{lowe1999object}.
In preliminary experiments, SIFT yielded lower accuracies while being less efficient to compute. Hence, it is not considered in our evaluation section.
% less efficient, worse results

% !TeX root = mi_attack_against_gan.tex
\section{Conclusion}
\label{sec:conclusion}

We suggest two membership inference attacks for generative models: the \textit{Monte Carlo (MC) attack} and the \textit{Reconstruction attack}. While the first is applicable to all generative models the latter is specialized for VAEs. Both attacks significantly outperform state of the art attacks against generative models often yielding accuracies close to $100\%$. In particular, the Reconstruction attack against VAEs outperformed all other attacks on all datasets. For CIFAR-10 the single and set MI even reached $\approx100\%.$
Even with dropout or more training data, the accuracies have proven robust.

On datasets with very good sample quality the MC attack outperformed state of the art. This supports the use of our formulated attacks to evaluate both overfitting and information leakage of generative models. On a dataset with very poor sample quality, however, the white box-attack \cite{hayes2017logan} outperformed our approaches. This is not very surprising as the MC attacks rely on a replication of training data characteristics which cannot be observed if the sample quality is insufficient.

%To systematically evaluate our attacks \blue{we formulated two actors: an honest-but-curious adversary $\mathcal{A}$ performing single MI, and a regulator $\mathcal{R}$ performing set MI}. Both previous work and novel attacks contributed in this paper are special cases of these types since only the function $\hat{f}(x)$ has to be specified. For the success of the attacks, it is crucial that this function attains higher values for training data records than for test data records. While single MI generalizes the setting of Hayes et al.~\cite{Hayes2019}, set MI yields higher accuracy values. Thus even slight information leakage can be recognized.

%High attack accuracies also suggest a weaker capability to generalize.
%Such cases of overfitting are easily detectable for discriminative models by the train-test-gap, but it is not trivial to assess the generalization behavior for generative models. There is a close relationship between overfitting and the success of membership inference and model inversion attacks. Hence, the attacks formulated within this work are well suited for quantitative evaluation of generative models. We see the attacks as part of a quality assessment process for trained models included in MLaaS to assist the user. Our attacks were able to succeed in cases where the white-box attack failed to recognize overfitting. Hence, they provide additional insights into the generalization behavior of the models which is still an open research issue. 

In general, we observed in this work that VAEs are more vulnerable to the MI attacks. This suggests that VAEs are more prone to overfitting than GANs if the same amount of training data is available. %This was also observed in the case of the AIS measurements by Wu et al.~\cite{wu2016quantitative}. 
Hence, the novel MI attacks formulated within this work give insights into the performance of different generative models and regularization techniques. In particular, the use of GANs being less vulnerable while producing detailed samples is motivated.

%With the increased use of generative models, the corresponding information leakage should be evaluated to avoid both overfitting and privacy issues.
%First results show that techniques to reduce overfitting, in particular regularization applied during the training process, or even the use of differential privacy can be successfully applied to generative models, too. However, they trade off increased training data protection for the quality of the results they produce.

%Note that it is possible to extend the applicability of the Reconstruction attack to other classes of generative models following the lines of the black-box attack~\cite{Hayes2019}: for this one would only need to train an auxiliary VAE on samples of the given generative model.
\section{Acknowledgements}
\label{sec:ack}
\blue{We thank the anonymous reviewers and our shepherd, Shruti Tople, for critically reading this paper and suggesting numerous improvements.
This work has received funding from the European Union’s Horizon 2020 research and innovation program under grant agreement No. 825333 (MOSAICROWN).}

\bibliographystyle{abbrv}
\bibliography{mi_attack_against_gan}

\onecolumn
%\onecolumngrid
\appendix
\section*{Appendix: Additional Figures}\label{app:figures}
\begin{figure*}[!h]
	\centering
	\subfloat[GAN on MNIST after 500 epochs]
	{
		\includegraphics[width=8cm]{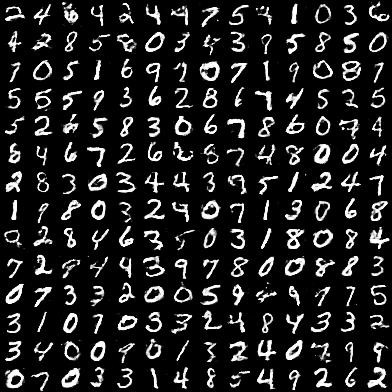}
	}
	\subfloat[VAE on MNIST after 300 epochs]
	{
		\includegraphics[width=8cm]{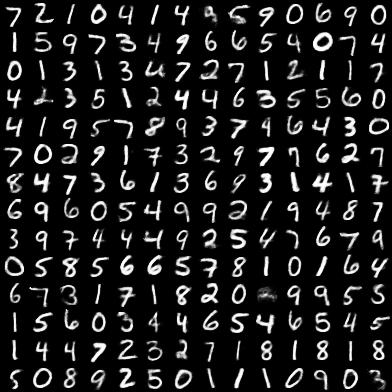}
	}
	\vfill
	\subfloat[VAE with 90\% Keep Probability]
	{
		\includegraphics[width=5.25cm]{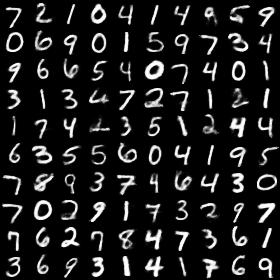}
	}
	\subfloat[VAE with 70\% Keep Probability]
	{
		\includegraphics[width=5.25cm]{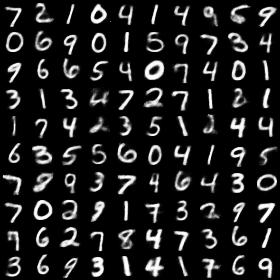}
	}
	\subfloat[VAE with 50\% Keep Probability]
	{
		\includegraphics[width=5.25cm]{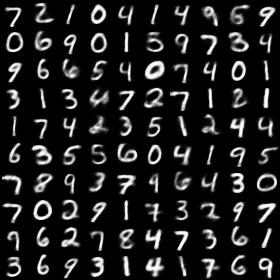}
	}
	\caption{Generated samples of the trained models.\\}
	\label{fig:generated_img_vae_gan_mnist}
\end{figure*}
\begin{figure*}[!h]
	\centering
	\subfloat[GAN on Fashion MNIST after 500 epochs]
	{
		\includegraphics[width=8cm]{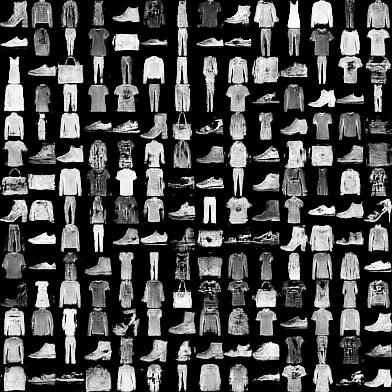}
	}
	\hspace{1em}
	\subfloat[VAE on Fashion MNIST after 300 epochs]
	{
		\includegraphics[width=8cm]{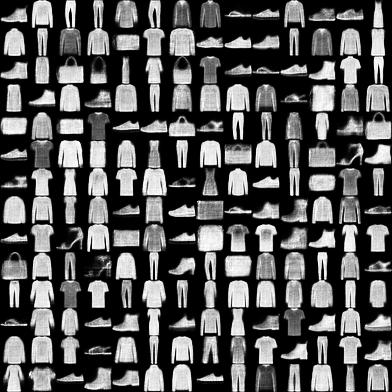}
	}
	\caption{Generated samples of the trained models.\\}
	\label{fig:generated_img_vae_gan_fashion}
\end{figure*}
\begin{figure*}[!h]
	\centering
	\subfloat[GAN after 1000 epochs]
	{
		\includegraphics[height=8cm]{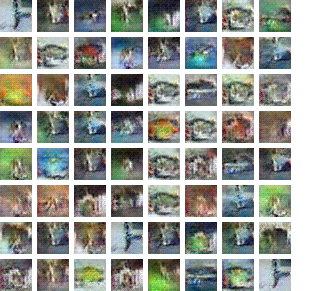}
	}
	\subfloat[VAE after 1200 epochs]
	{
		\includegraphics[height=8cm]{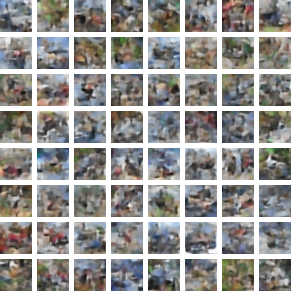}
	}
	\caption{Generated images of a GAN and a VAE after training on CIFAR-10 dataset.}
	\label{fig:generated_cifar_vae_gan}
\end{figure*}
\end{document}